\documentclass[referee]{aa}
\usepackage[varg]{txfonts}
\usepackage[utf8]{inputenc}
\usepackage[english]{babel}
\usepackage{indentfirst}
\usepackage{graphicx}
\usepackage[]{natbib}
\bibpunct{(}{)}{;}{a}{}{,} 
\usepackage{lineno}

\newcommand{\rem}[1]{}

\begin{document}

\title{New data support the existence of the Hercules-Corona Borealis Great Wall  }
 
\author{Istv\'an Horv\'ath\inst{1}
  \and Zsolt Bagoly\inst{2}  
  \and Jon Hakkila\inst{3}
  \and L. Viktor Tóth\inst{2}  
  }

\offprints{I. Horv\'ath, \email{horvath.istvan@uni-nke.hu}}

\institute{National University of Public Service, Budapest, Hungary,
  \and E\"otv\"os  University, Budapest, Hungary
  \and University of Charleston South Carolina at the College of Charleston, Charleston, SC, USA,
  }

\date{Received  August, 2014 / Accepted ber, 2015}

\abstract
{
Several large structures, including the Sloan Great Wall, 
the Huge Large Quasar Group, and a large gamma-ray burst cluster
referred to as the Hercules-Corona Borealis Great Wall, 
appear to exceed the
maximum structural size predicted by Universal inflationary models.
The existence of very large structures such as these might necessitate
cosmological model modifications.
}
{Gamma-ray bursts are the most luminous sources found in nature.
They are associated with the stellar endpoints of massive stars and are 
found in and near distant galaxies. Since they are viable indicators 
of the dense part of the Universe containing normal matter, the spatial 
distribution of gamma-ray bursts can serve as tracers
of Universal large-scale structure.} 
{ 
An increased sample size of gamma-ray bursts with known redshift
provides us with the opportunity to validate or invalidate the existence of
the Hercules-Corona Borealis Great Wall. Nearest-neighbour 
tests are used to search the larger sample for evidence of clustering
and a bootstrap point-radius method is used to estimate the angular
cluster size. The potential influence of angular sampling biasing is
studied to determine the viability of the results.  
}
{The larger gamma-ray burst database further supports the existence
of a statistically significant gamma-ray burst cluster at $1.6 \le z < 2.1$
with an estimated angular size of 2000-3000 Mpc.
} 
{
Although small number statistics limit our angular resolution and
do not rule out the existence of adjacent and/or line-of-sight smaller structures,
these structures must still  clump together in order for
us to see the large gamma-ray burst cluster detected here. This cluster
provides support for the existence of very large-scale universal heterogeneities.
}

\maketitle 

\section{Introduction}\label{sec:intro}

The high luminosities of gamma-ray bursts (GRBs) make them ideal
candidates for probing large-scale Universal structure. 
Gamma-ray bursts signify the presence of stellar endpoints and thus trace 
the location of matter in the universe. This is true whether they are long bursts 
(presumably originating from hypernovae), short bursts 
(presumably originating from compact objects), or intermediate bursts 
(with unknown origins that are still likely related to stellar endpoints).
Assuming that the Universe is homogeneous and isotropic
on a large scale implies that the large-scale distribution of GRBs 
should similarly be homogeneous and isotropic. The angular isotropy of GRBs has been
well-studied over the past few decades \citep{Briggs96,bal98,bal99,mesz00,mgc03,vbh08}.
For the most part, GRBs are distributed uniformly, although some subsamples
(generally believed to be those with lower luminosities and therefore thought
to be cosmologically local) appear to deviate from isotropy
\citep{bal98,Cline99,mesz00,li01,mgc03,vbh08}. We have recently identified
a surprisingly large anisotropy in the overall GRB angular distribution, 
suggestive of clustering, at redshift two in the constellations of Hercules
and Corona Borealis. The underlying distribution of
matter suggested by this cluster is large enough to question
standard assumptions about the largest scale of Universal structures.

We revisit the angular and radial distributions of GRBs
with known redshifts in an attempt to reexamine our previous 
claims suggesting the existence of this structure. 
As of November 2013, the redshifts of 361 GRBs have been 
determined\footnote{http://lyra.berkeley.edu/grbox/grbox.php or http://www.astro.caltech.edu/grbox/grbox.php}; this represents
an increase in sample size of 28\% over that used in our
previous analysis (283 bursts observed up until July 2012). 
The number of GRBs in the $z=2$ redshift range,
where the cluster resides, has increased from 31 bursts
to 44 bursts, a 42\% sample size increase that is large enough
to warrant an updated analysis.
We apply k$^{\text{th}}$ nearest neighbour analysis and the
bootstrap point radius method to this database composed largely
of bursts detected by NASA's Swift experiment.

\section{Nearest-neighbour statistics: the largest anisotropy
is at $z=2$}\label{sec:GRB spatial distribution}

The larger GRB database allows us to re-examine the significance of 
our prior results. We employ the same statistical tests so as to 
retain consistency in our methodology and not
introduce any potential analysis biases.

The GRB sample is subdivided by redshift $z$ in a manner
similar to our previous work \citep{hhb14} 
so that we can base our angular studies on well-defined distance 
groupings. The GRB redshift uncertainties are small (many 
GRB redshifts are quoted to three or four significant figures), so
it is possible in theory to create a large number of radial groups or bins
and thus maintain a small $z$-dispersion in the sample.
The drawback to this approach is that the 361 burst
sample is still small, and angular resolution is limited based on
the number of bursts in each radial group. 
We have subdivided the total sample into eight separate cases, containing 
the following numbers of radial
groups: two, three, four, five, six, seven, eight, and nine. 
These choices allow us to examine bulk anisotropies in the GRB
distribution over various distance ranges. 
However, binning the data limits the angular
resolution we can realistically obtain within each radial bin: we are
capable of finding large anisotropies.
These cases are not independent of one another; each 
contains the same GRB sample binned differently.
A choice of
one radial group corresponds to the bulk angular distribution of GRBs
in the plane of the sky; we do not analyse  group one here, since
it does not make use of the carefully measured redshifts we employed. In contrast, the choice of nine radial groups provides us with radial bins 
with the smallest number of bursts per bin ($\approx 40$) for
which  we can make reasonable, quantifiable estimates on bulk
anisotropies. 
When choosing between 2 and 9 radial divisions, we keep the
numbers of bursts in each radial group identical. The result of
this approach is that we exclude GRBs with the smallest
redshifts in some cases. 
For example, in the four group case the closest GRB (with the
smallest redshift) was excluded, therefore,
each of the four groups
contains 90 GRBs (361=4x90+1).

\begin{figure}[h!]\begin{center}
  \resizebox{.95\hsize}{!}{\includegraphics[angle=0]{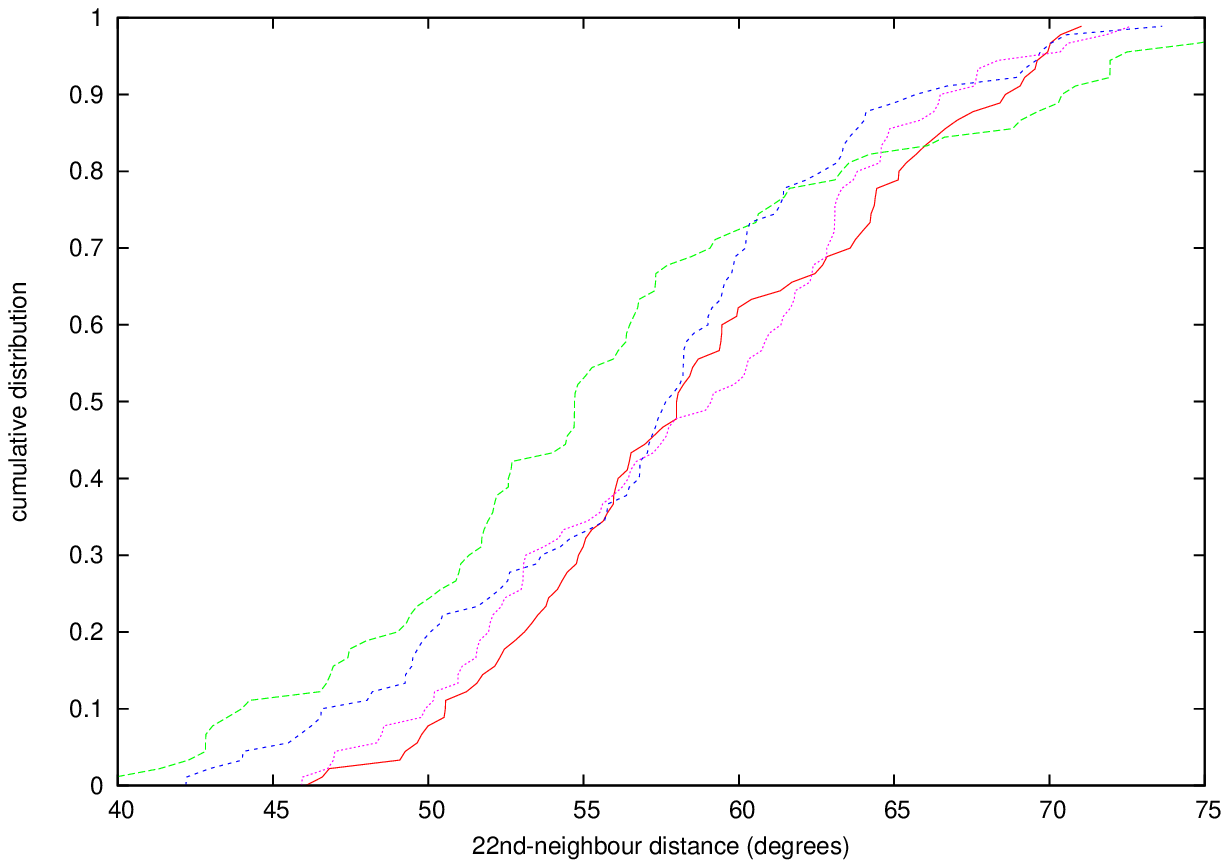}}
  \resizebox{.95\hsize}{!}{\includegraphics[angle=0]{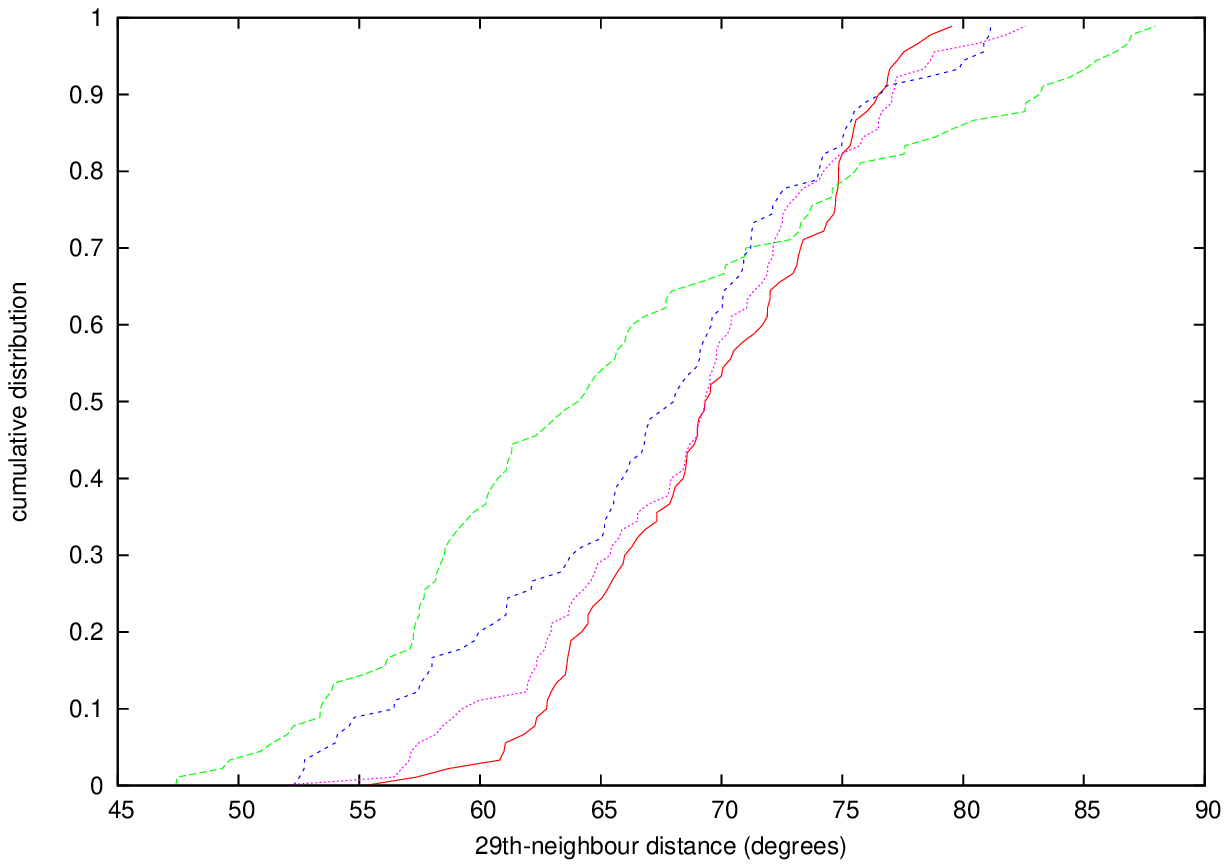}  }
  \caption{\small{
                The 22nd (top) and 29th (bottom) neighbour distribution for the four group
case (each group contains 90 GRBs), red, green,
blue, and pink identifies radial groups 1, 2, 3, and 4.
  }}
  \label{fig:0906}
\end{center}
\end{figure}

We examine the angular burst distributions of each radial group
by independently applying the kth nearest-neighbour statistics
to the bursts in each group. As an example of our procedure, 
we consider the four group case again.  For each radial group, 
we calculate the angular separations between all 90 GRBs.
All neighbours of each GRB are identified and ranked as the nearest, second nearest, etc. 
The 90 nearest neighbour separations are collected into a first distribution,
the 90 second-nearest neighbour separations are collected into a second distribution,
and the process is repeated with each set of neighbours until the orthogonal set is 
completed with the 89th distribution being composed of the 89th nearest
neighbour (farthest) separations.
For each group in the four group case, these 89 nearest-neighbour distributions
can be compared across the groups using a Kolmogorov-Smirnoff test.
As this has been accomplished for the four group case, the same approach can be 
applied to the available nearest-neighbour distributions for all eight radial 
groupings (the two group case through the nine group case).

Each of the eight radial groupings indicates that significant anisotropies are present 
in one specific radial region, as defined by redshift.
In other words, most of the kth nearest-neighbour distributions are
not significantly different, except those that are close to one specific redshift.
The bin containing the largest cluster of GRBs always comes from
the redshift range $1.6 \le z < 2.1,$ as  found in our previous work \citep{hhb13,hhb14}.
Figure 1 shows an example of when the sample is divided into four radial groups.
In this case, each group contains 90 GRBs in the redshift ranges
$2.68 \le z < 9.4$ (group 1), $1.61 \le z < 2.68$ (group 2), $0.85 \le z < 1.61$ (group 3), and 
$0 \le z < 0.85$ (group 4).
For this example, Table 1 shows the probability that the two distributions 
are different. Boldface type indicates that the significance of the  31st 
nearest-neighbour distributions of two groups are different
by more than 3$\sigma $. There are no significant differences within 
group 1, group 3, and group 4 distributions, but the 31st 
nearest-neighbour distributions in group 2 are significantly different from the other distribution.
The 31st nearest-neighbour distribution is just an example demonstrating a group 2 
anisotropy; the same is also true for the 22nd, 23rd, ... 55th nearest-neighbour distributions.
The GRBs clustering on small angular scales would show differences when
describing close neighbour pairs, while GRBs found on opposite sides of the celestial
sphere would exhibit differences when describing distant neighbour pairs.
The GRBs in the $1.61 \le z < 2.68$ redshift range have a preference for neighbours
with moderately close angular separations, suggesting a large angular cluster.

\begin{table}[t]\begin{center}
    \hfill{}
    \caption{Results of the 31st nearest-neighbour distributions comparing the GRB groups 
    (in the four group case). 
   Comparing the distributions of two groups for the 31st nearest-neighbour, the 
   numbers in this table are the significance of the null hypothesis that the two distributions 
   are different. Boldface type indicates that there 
    are significant (more than 3$\sigma $) differences between the two groups.}
        \begin{tabular}{|l||c|c|c|c|c|c|c|c|c|}\hline 
                 & $z_{min}$ &  gr2  & gr3 & gr4  \\ \hline \hline 
gr1 & 2.68 &  {\bf 0.9999999}  & 0.942 & 0.672  \\ \hline
gr2 & 1.61 &    & {\bf 0.99904}  & {\bf 0.9999988}   \\ \hline
gr3 & 0.85 &    &  & 0.960   \\ \hline
        \end{tabular}
        \hfill{}
        \label{tab:err}
\end{center}
\end{table}

\section{Bootstrap point-radius method: the anisotropy represents a large GRB cluster}

As demonstrated in the previous section, nearest-neighbour tests
identify pairing consistent with a large, loose GRB cluster in the redshift range
$1.6 < z \le 2.1$.
The significance of this cluster can also be measured using
other statistical tests designed to identify clustering.
Among these is the bootstrap point-radius method
described in Sect. 5 of \citet{hhb14}.
The updated data set to which we apply this test
contains 44 GRB in the redshift interval 
$1.6 < z \leq 2.1$.

Our use of the bootstrap point-radius method
assumes that the sky exposure is independent of $z$.
To carry out our analysis, we choose 44 GRBs from the observed
data set and compare the sky distribution of
this subsample with the sky distribution
of 44 GRBs with $1.6 < z \leq 2.1$

To study the selected bursts in two dimensions,
we select random locations on the
celestial sphere and find how many of the
44 points lie within a circle of predefined angular 
radius, for example, within $20\,^{\circ}$. 
We build statistics for this test
by repeating the process a large number of times (i.e., 10000).
From the 10000 Monte Carlo runs, we select
the largest number of bursts found within the 
angular circle.

This analysis can be performed with the clustered
44 GRB positions as well as with 44 randomly chosen 
GRB locations from the observed data. 
There are some angular radii for which
the maximum with the 44 GRBs with $1.6 < z \leq 2.1$ is
significant. We repeat the process with 44 different randomly
chosen burst positions, and we repeated the experiment 17500 times 
 to understand the statistical variations of this subsample.
We also perform the same measurement using angular circles
of different radii. 
The frequencies obtained are shown in Fig. 2.

\begin{figure}[h!]\begin{center}
 
 \resizebox{\hsize}{!}{\includegraphics[angle=0]{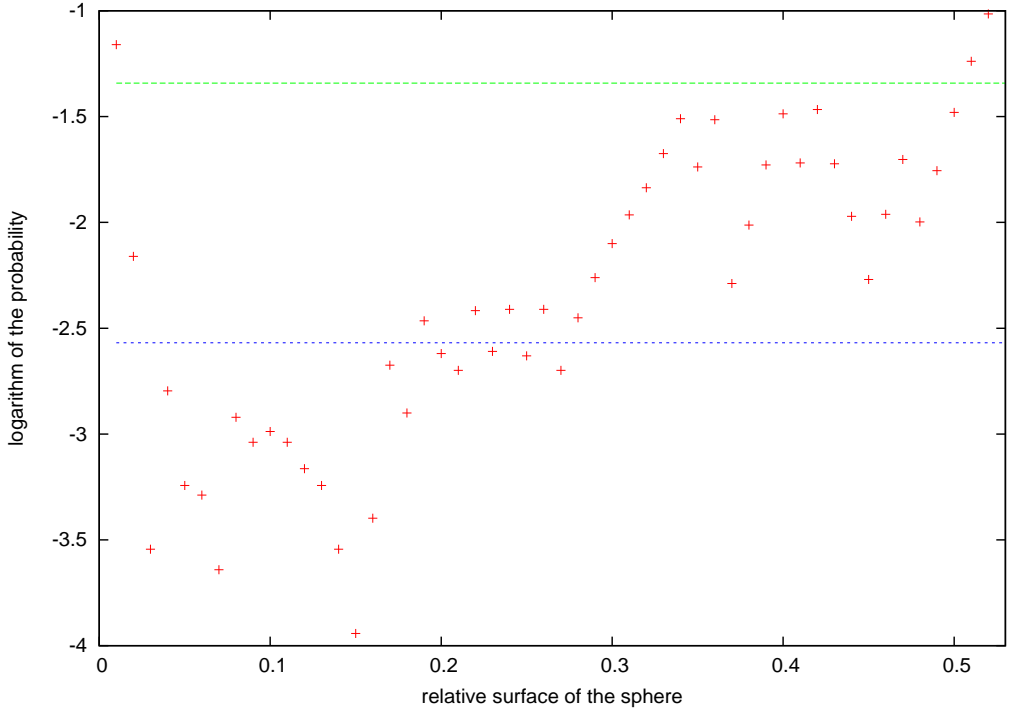} }
 
 \caption{\small{Results of the Monte-Carlo bootstrap point-radius method. 
The horizontal coordinate is the area of the circle in the sky
relative to the whole sky ($4\pi $). The vertical
coordinate is the logarithm of the frequency 
found from the 17500 runs. Green (blue) line shows the $2\sigma $ 
($3\sigma $) deviations.}}
  \label{fig:0906fd}
\end{center}
\end{figure}

Figure 2 demonstrates that the 9-18 \% of the
sky identified for $1.6 < z \leq 2.1$ contains significantly
more GRBs than similar circles at other GRB redshifts.
When the area is chosen to be 0.0375 $\times 4 \pi$ (corresponding
to an angular radius of $\theta_{\rm max} = 22.3 ^\circ$), 
13 out of the 44 GRBs lie inside the circle.
When the area is chosen to be 0.0875 $\times 4 \pi$
($\theta_{\rm max} = 34.4 ^\circ$),
18 of 44 GRBs lie inside the circle. 
When the area is chosen to be 0.1875 $\times 4 \pi$
($\theta_{\rm max} = 51.3 ^\circ$),
25 GRBs out of the
44 lie inside the circle. In this last case, only two out
of the 17500 bootstrap cases had 25 or more GRBs inside the circle. 
This result is, therefore, a statistically significant (p=0.0001143) deviation, and the binomial
probability for this being random is  $p_b=2\times 10^{-8}$.

The $42\%$ increase in sample size should have led to 
a noticeable decrease in significance if the sample represented random sampling.
However, in the radii between roughly $4^\circ$ and $90^\circ$,
49 angular circles contain enough GRBs to
exceed the $2\sigma $ level, compared to 28 found in our previous analysis \citep{hhb14}).
Additionally, there are 16 angular circles containing enough
GRBs to exceed the $3\sigma$ level (compared to only two in our
previously published result),
therefore, the evidence has strengthened that these bursts are 
mapping out some large-scale universal structure.

\section{Sky exposure: Sampling biases do not appear to be responsible for the anisotropy}

Observing biases can introduce measurable angular anisotropies in a sample. 
However, prior results suggest that these biases
are unlikely to be responsible for the observed cluster
at $z \approx 2$.
The largest potential causes of 
angular biasing are:
\begin{itemize}
\item Sky exposure. This is a well-known bias describing
favoured detection of GRBs in some angular directions over
others. Sky exposure is a function of instrumental response rather
than a true source distributional preference; some causes
of anisotropic sky exposure include spacecraft pointing and
a preferred orbital plane, the avoidance of certain pointing
directions such as the Sun or occultation by the Earth.
\item Anisotropic measurement of GRB redshifts.
GRB redshift measurements are made in the visual/infrared
by ground-based telescopes, and are thus affected
by observatory latitudes, seasonal weather, and
Galactic extinction.
\end{itemize}

Each GRB instrument samples the sky differently, 
making the summed sky exposure difficult to identify for our 
heterogeneous GRB sample, which has been observed by 
many instruments since the late 1990s. 
However, since more than $3/4$ of our sample was detected by Swift, 
the Swift sky exposure dominates the sampling.
Thus, we assume to first order that Swift's sky exposure is a reasonable
approximation of the sky exposure of the entire burst sample. 
Because of its orbital characteristics, Swift \citep{Bau13} has sampled
ecliptic polar regions at slightly higher rates than ecliptic equatorial regions.
Our simple model assumes that ecliptic polar regions
($| \beta | \ge45\,^{\circ}$, where $\beta$ is the ecliptic latitude) 
are sampled $1.83$ times more frequently than
the ecliptic equatorial region.

The location of ground-based optical and infrared telescopes measuring GRB
redshifts can also lead to anisotropic observations. However,
since a large number of ground-based telescopes at a variety of latitudes
and longitudes have been used in GRB follow-up observations, there does
not appear to be an Earth-based bias that would favour GRB afterglow
measurements in some sky locations over others. Thus, our 
sampling model does not include a term accounting for telescope location.

Extinction due to dust from the Milky Way disk does not affect the 
detection of GRBs, but it does affect redshift measurements
in an angularly-dependent way. 
Extinction removes light from
extragalactic sources, making it harder to measure spectral
characteristics from which redshifts can be obtained.
Although the Galactic dust is strongly 
concentrated towards the
Galactic equator, it is also very clumpy. This clumpiness
makes the effect of extinction on measuring GRB redshifts very difficult to model; 
the details of the process depends on many variables, such as 
the Galactic latitude and longitude of the burst, the intrinsic luminosity 
and decay rate of the afterglow, the light-gathering ability of the telescope and 
the instrumental response of the spectrograph used, the redshift of the burst, 
and the observing conditions at 
the time of detection. 

We check to see whether or not the GRB sample favours
low-extinction regions by examining the distribution of 
visual extinctions in the directions of these 361 bursts. 
Extinctions are obtained from the 
high angular resolution DIRBE catalogue of \citealt{sch11}
found online at  http://irsa.ipac.caltech.edu/applications/DUST/. 
The results, shown in Fig. 3, are that the sample can
be modelled by a lognormal distribution centred at $A_v=0.13$ mag
with standard deviations $\sigma^+=0.22$ and $\sigma^-=0.08$.
Fully $91\%$ of the bursts in the sample have visual extinctions of
$A_v \le 0.5$ mag, indicating that a characteristic of a GRB with a
measured redshift is that it is not obscured by Galactic extinction.

\begin{figure}[h!]\begin{center}
 
 \resizebox{\hsize}{!}{\includegraphics[angle=0]{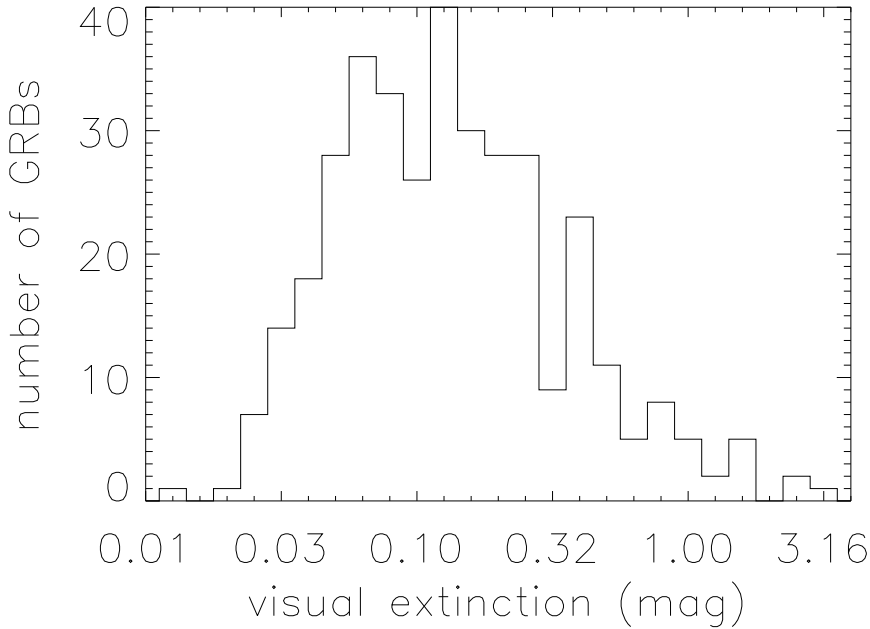} }
 
 \caption{\small{Galactic visual extinctions in the directions of the 361 GRBs 
 in this sample. The measurement of GRB redshifts
 strongly favours small amounts of interstellar extinction.}}
  \label{fig:0906fd}
\end{center}
\end{figure}

It is not possible to tell if, in trying to measure 
GRB redshifts from afterglows, observers have avoided
pointing their optical telescopes in the direction of GRBs that lie
too close to the Galactic equator. This kind of a bias is possible
given the potential low return on afterglow detection
(especially for small- and medium-sized telescopes),
but this bias cannot be modelled with the limited information 
provided by optical observers.
Since the development of a model describing
extinction and extinction-related biases on redshift measurement
is very difficult, to estimate the effects of extinction 
on the sample we use all GRBs with measured redshifts
found within $20\,^{\circ}$ of the Galactic equator relative to all
GRBs with measured redshifts. We find that only $3.1\%$ of the GRB sample
has had redshift measurements made within $20\,^{\circ}$ of the Galactic equator.

The Galactic equatorial region is the poorest-sampled region.
The ecliptic polar regions ($b \ge 20\,^{\circ}$ and $| \beta | \ge45\,^{\circ}$
are the best-sampled regions and the ecliptic equatorial regions
($b \ge 20\,^{\circ}$ and $| \beta | < 45\,^{\circ}$) are well-sampled regions.
Correcting the observations for this biasing, the expected numbers of 
GRBs in each group are 
$7.50$ (best, north), $7.50$ (best, south), $11.2$ (good, north), $11.2$ (good, south), 
and $5.54$ (poor), whereas the actual counts are
$13$ (best, north), $2$ (best, south), $12$ (good, north), $8$ (good, south), 
and $7$ (poor).
This results in a $\chi^2$ probability of $p=0.051$
that this is due to chance. 

Although this probability is higher than that quoted previously ($p=0.025$),
the cluster density has increased relative to the rest of the $z \approx 2$ sky since
the last analysis.
Many of the new bursts have been detected just outside the edge of the 
best-sampled region, in the northern well-sampled region. Unfortunately,
our low-resolution angular bias correction treats all GRBs in the well-sampled
region as if they are not part of the cluster.  If we naively assume that the cluster
comprises 17 observed bursts (13 in $50\%$ of the northern best-sampled 
region and 4 in $10\%$ of the northern well-sampled region) and recalculate 
the probability that this clustering is random, the probability changes to 
$p = 1.6 \times 10^{-4}$ that exposure is responsible for the clustering. 
This calculation also suggests that the cluster properties might be 
affected slightly by exposure: the few bursts seen in regions with less 
exposure could represent a larger number of undetected bursts.
The cluster might be shifted several degrees west of where
we have previously identified it. 

\section{Summary and conclusion}

The evidence for a possible large-scale Universal 
structure \citep{hhb14} at a redshift of $z\approx2$
has strengthened,
using a larger database of GRBs with known redshift. The new sample
contains $28\%$ more bursts than the previous sample, and $42\%$ more
bursts in the $1.6 \le z < 2.1$ redshift range.
Because the cluster has become more populated relative to the rest of the
angular distribution at the same redshift, 
our angular tests have returned more significant results.
Nearest-neighbour tests indicate that GRBs in this redshift range
favour each others' presence through moderate angular separations.
The two-dimensional bootstrap point-radius method reaches the $3\sigma$ 
level for a number of different angular radii, indicating a large GRB cluster. 
Although sampling biases are present and are significant, these
coupled with small number statistics, do not seem to be responsible for the observed
clustering of GRBs at this redshift.

\begin{figure}[h!]\begin{center}
  \resizebox{.95\hsize}{!}{\includegraphics[angle=270]{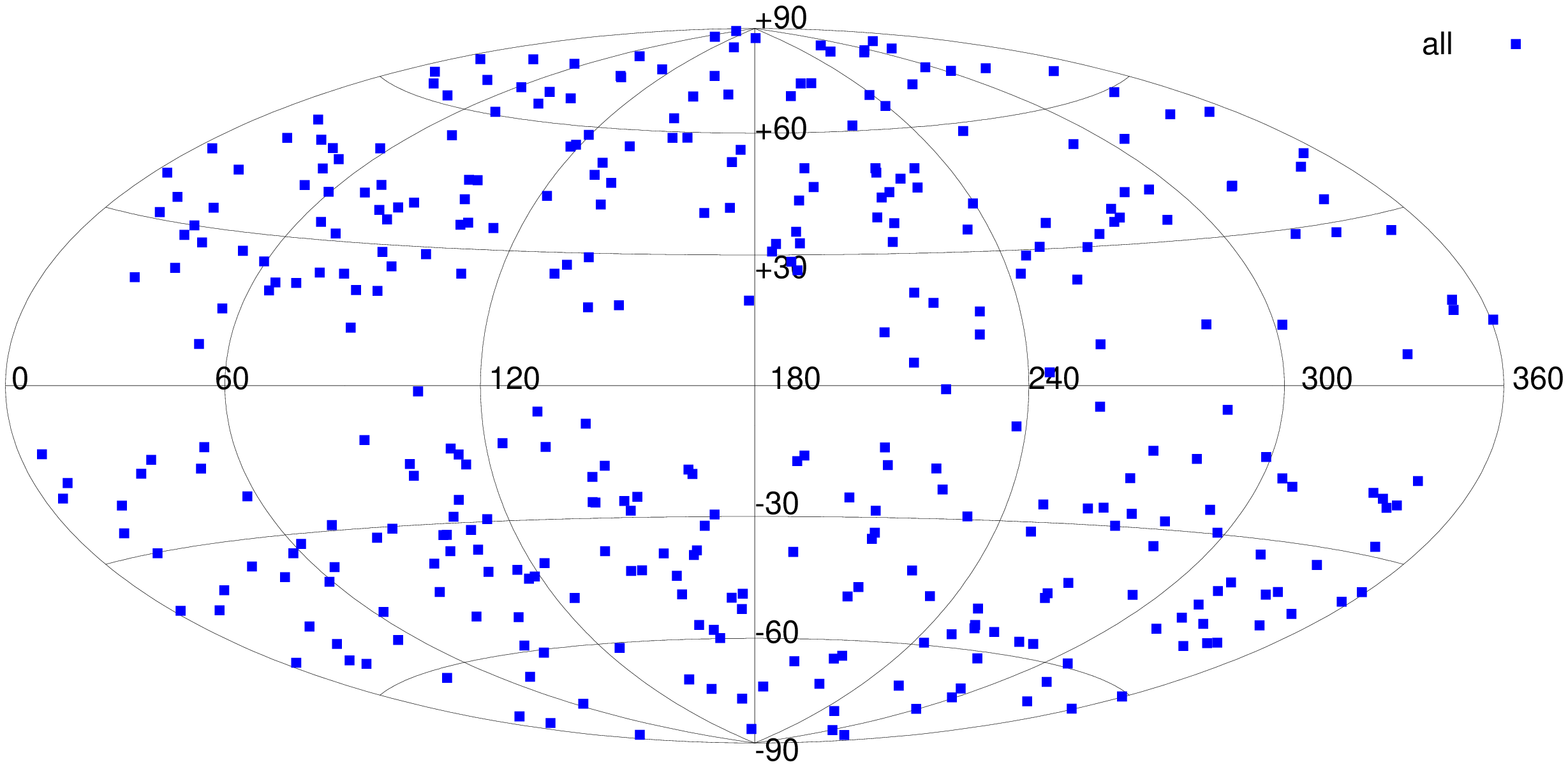}}
  \resizebox{.95\hsize}{!}{\includegraphics[angle=270]{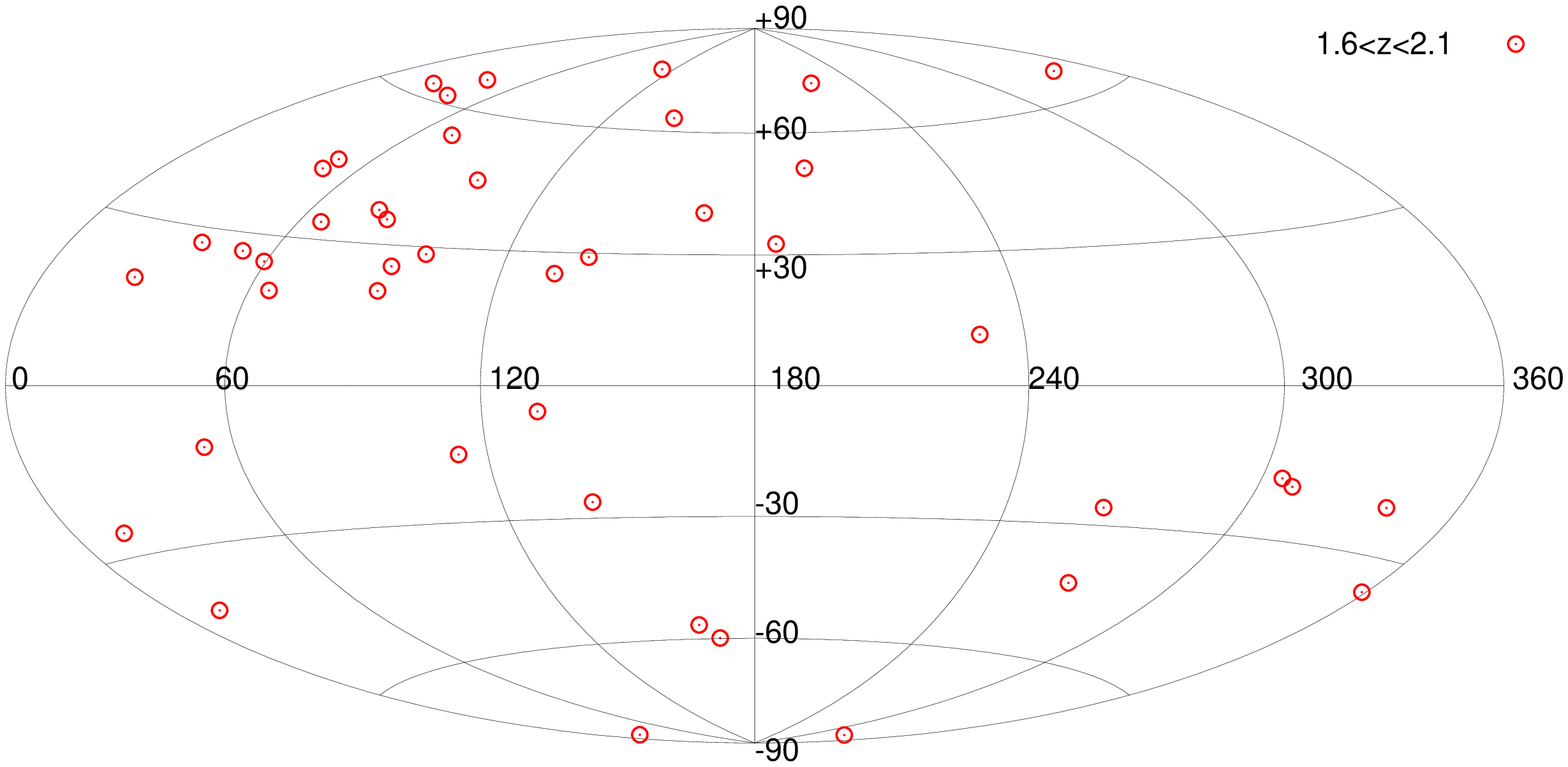}  }
  \caption{\small{
                \textsl{Top:}  distribution of GRBs with measured redshift
                (blue). Although the distribution of all GRBs is
                fairly isotropic, extinction causes this sample to miss GRBs
                near the Galactic plane.
        \textsl{Bottom:}  anisotropic distribution of GRBs near redshift $z=2$ (red).
  }}
  \label{fig:0906}
\end{center}
\end{figure}

GRBs are the most luminous known form of energy release
available to normal matter. As such, they
are tracers for the presence of normal matter that can be
detected at distances where the matter is otherwise too 
faint to be observed. The GRB cluster at $z \approx 2$
appears to identify the presence of a larger angular
structure that covers almost one-eighth of the sky.
This encompasses half of the constellations of Bootes, 
Draco, and Lyra, and
all of the constellations of Hercules and Corona Borealis.
This structure has been given the popular name of the
{\em \textup{Hercules-Corona Borealis Great 
Wall}}, or Her-CrB GW.

We estimate the size of the Her-CrB GW to be about 2000-3000 Mpc
across. Few limits on its radial thickness exist, other than because it
appears to be confined to the $1.6 \le z < 2.1$ redshift range.
This large size makes the structure inconsistent with current
inflationary Universal models because 
it is larger than the roughly 100 Mpc limit thought to signify
the End of Greatness at which large-scale structure ceases.

However, the Her-CrB GW is not the first optical/infrared structure found
to exceed the 100 Mpc size limit.
Several large filamentary structures have been identified using
optical and infrared redshifts of galaxies; these include the
200 Mpc-sized CfA2 Great Wall \citep{Gel89} and the
400 Mpc-sized Sloan Great Wall \citep{Gott05}. 
In the ensuing years, other structures
have been identified using quasars; the largest of these
is the Huge Large Quasar Group (Huge-LQG) \cite{clo12},
which has a length of more than 1400 Mpc. Most recently,
\cite{sza15}  found a 440 Mpc diameter supervoid aligned
with a cold spot on the cosmic microwave background.

On the other hand, many results support the  cosmological principle. 
\cite{Yah2005} reported 
that the galaxy distribution was homogeneous on scales larger than
$60-70 \, h^{-1}$ Mpc. \cite{Bag2008}
showed that the fractal dimension makes
a rapid transition close to 3 at 40 -
100 Mpc scales. \cite{Sar2009} found the galaxy distribution to be
homogeneous at length-scales
greater than $70 \, h^{-1}$ Mpc, and
\cite{Yad2010} estimated the homogeneity upper limit scale was
close to $260 \, h^{-1}$ Mpc.

As large as it appears to be, the Her-CrB GW does not necessarily
have to violate the basic assumptions of the cosmological principle
(the assumptions of a homogeneous and isotropic Universe).
Theoretical large-scale structure models indicate
that some structures will exceed the End of Greatness on
purely statistical grounds \citep{nada13}, and this may be one such structure
(albeit a very large one).
Along these lines, this may not be a single structure, but a
clustering of smaller adjacent and/or line-of-sight structures; the small number of bursts currently found in the cluster limits 
our ability to angularly resolve it.
However, this becomes a semantic issue at some point,
since a cluster of smaller structures is still a larger structure.

\begin{acknowledgements} 
This research was supported by OTKA grant NN111016 and by NASA EPSCoR grant NNX13AD28A. 
We thank the anonymous referee for comments that greatly improved this paper.
\end{acknowledgements}

\bibliographystyle{aa} 
\bibliography{horv14}

\hyphenation{Post-Script Sprin-ger}
\begin{thebibliography}{21}
\expandafter\ifx\csname natexlab\endcsname\relax\def\natexlab#1{#1}\fi

\bibitem[{{Bagla} {et~al.}(2008){Bagla}, {Yadav}, \& {Seshadri}}]{Bag2008}
{Bagla}, J.~S., {Yadav}, J., \& {Seshadri}, T.~R. 2008, \mnras, 390, 829

\bibitem[{{Bal{\'a}zs} {et~al.}(1998){Bal{\'a}zs}, {M{\'e}sz{\'a}ros}, \&
  {Horv{\'a}th}}]{bal98}
{Bal{\'a}zs}, L.~G., {M{\'e}sz{\'a}ros}, A., \& {Horv{\'a}th}, I. 1998, \aap,
  339, 1

\bibitem[{{Bal{\'a}zs} {et~al.}(1999){Bal{\'a}zs}, {M{\'e}sz{\'a}ros},
  {Horv{\'a}th}, \& {Vavrek}}]{bal99}
{Bal{\'a}zs}, L.~G., {M{\'e}sz{\'a}ros}, A., {Horv{\'a}th}, I., \& {Vavrek}, R.
  1999, \aaps, 138, 417

\bibitem[{{Baumgartner} {et~al.}(2013){Baumgartner}, {Tueller}, {Markwardt},
  {Skinner}, {Barthelmy}, {Mushotzky}, {Evans}, \& {Gehrels}}]{Bau13}
{Baumgartner}, W.~H., {Tueller}, J., {Markwardt}, C.~B., {et~al.} 2013, \apjs,
  207, 19

\bibitem[{{Briggs} {et~al.}(1996){Briggs}, {Paciesas}, {Pendleton}, {Meegan},
  {Fishman}, {Horack}, {Brock}, {Kouveliotou}, {Hartmann}, \&
  {Hakkila}}]{Briggs96}
{Briggs}, M.~S., {Paciesas}, W.~S., {Pendleton}, G.~N., {et~al.} 1996, \apj,
  459, 40

\bibitem[{{Cline} {et~al.}(1999){Cline}, {Matthey}, \& {Otwinowski}}]{Cline99}
{Cline}, D.~B., {Matthey}, C., \& {Otwinowski}, S. 1999, \apj, 527, 827

\bibitem[{{Clowes} {et~al.}(2013){Clowes}, {Harris}, {Raghunathan},
  {Campusano}, {S{\"o}chting}, \& {Graham}}]{clo12}
{Clowes}, R.~G., {Harris}, K.~A., {Raghunathan}, S., {et~al.} 2013, \mnras,
  429, 2910

\bibitem[{{Geller} \& {Huchra}(1989)}]{Gel89}
{Geller}, M.~J. \& {Huchra}, J.~P. 1989, Science, 246, 897

\bibitem[{{Gott} {et~al.}(2005){Gott}, {Juri{\'c}}, {Schlegel}, {Hoyle},
  {Vogeley}, {Tegmark}, {Bahcall}, \& {Brinkmann}}]{Gott05}
{Gott}, III, J.~R., {Juri{\'c}}, M., {Schlegel}, D., {et~al.} 2005, \apj, 624,
  463

\bibitem[{{Horv{\'a}th} {et~al.}(2013){Horv{\'a}th}, {Hakkila}, \&
  {Bagoly}}]{hhb13}
{Horv{\'a}th}, I., {Hakkila}, J., \& {Bagoly}, Z. 2013, 7th Huntsville
  Gamma-Ray Burst Symposium, GRB 2013: paper 33 in eConf Proceedings C1304143

\bibitem[{{Horv{\'a}th} {et~al.}(2014){Horv{\'a}th}, {Hakkila}, \&
  {Bagoly}}]{hhb14}
{Horv{\'a}th}, I., {Hakkila}, J., \& {Bagoly}, Z. 2014, \aap, 561, L12

\bibitem[{{Litvin} {et~al.}(2001){Litvin}, {Matveev}, {Mamedov}, \&
  {Orlov}}]{li01}
{Litvin}, V.~F., {Matveev}, S.~A., {Mamedov}, S.~V., \& {Orlov}, V.~V. 2001,
  Astronomy Letters, 27, 416

\bibitem[{{Magliocchetti} {et~al.}(2003){Magliocchetti}, {Ghirlanda}, \&
  {Celotti}}]{mgc03}
{Magliocchetti}, M., {Ghirlanda}, G., \& {Celotti}, A. 2003, \mnras, 343, 255

\bibitem[{{M{\'e}sz{\'a}ros} {et~al.}(2000){M{\'e}sz{\'a}ros}, {Bagoly},
  {Horv{\'a}th}, {Bal{\'a}zs}, \& {Vavrek}}]{mesz00}
{M{\'e}sz{\'a}ros}, A., {Bagoly}, Z., {Horv{\'a}th}, I., {Bal{\'a}zs}, L.~G.,
  \& {Vavrek}, R. 2000, \apj, 539, 98

\bibitem[{{Nadathur}(2013)}]{nada13}
{Nadathur}, S. 2013, \mnras, 434, 398

\bibitem[{{Sarkar} {et~al.}(2009){Sarkar}, {Yadav}, {Pandey}, \&
  {Bharadwaj}}]{Sar2009}
{Sarkar}, P., {Yadav}, J., {Pandey}, B., \& {Bharadwaj}, S. 2009, \mnras, 399,
  L128

\bibitem[{{Schlafly} \& {Finkbeiner}(2011)}]{sch11}
{Schlafly}, E.~F. \& {Finkbeiner}, D.~P. 2011, \apj, 737, 103

\bibitem[{{Szapudi} {et~al.}(2015){Szapudi}, {Kov{\'a}cs}, {Granett}, {Frei},
  {Silk}, {Burgett}, {Cole}, {Draper}, {Farrow}, {Kaiser}, {Magnier},
  {Metcalfe}, {Morgan}, {Price}, {Tonry}, \& {Wainscoat}}]{sza15}
{Szapudi}, I., {Kov{\'a}cs}, A., {Granett}, B.~R., {et~al.} 2015, \mnras, 450,
  288

\bibitem[{{Vavrek} {et~al.}(2008){Vavrek}, {Bal{\'a}zs}, {M{\'e}sz{\'a}ros},
  {Horv{\'a}th}, \& {Bagoly}}]{vbh08}
{Vavrek}, R., {Bal{\'a}zs}, L.~G., {M{\'e}sz{\'a}ros}, A., {Horv{\'a}th}, I.,
  \& {Bagoly}, Z. 2008, \mnras, 391, 1741

\bibitem[{{Yadav} {et~al.}(2010){Yadav}, {Bagla}, \& {Khandai}}]{Yad2010}
{Yadav}, J.~K., {Bagla}, J.~S., \& {Khandai}, N. 2010, \mnras, 405, 2009

\bibitem[{{Yahata} {et~al.}(2005){Yahata}, {Suto}, {Kayo}, {Matsubara},
  {Connolly}, {vanden Berk}, {Sheth}, {Szapudi}, {Anderson}, {Bahcall},
  {Brinkmann}, {Csabai}, {Fan}, {Loveday}, {Szalay}, \& {York}}]{Yah2005}
{Yahata}, K., {Suto}, Y., {Kayo}, I., {et~al.} 2005, \pasj, 57, 529

\end{thebibliography}

\section{Appendix}

\begin{table}[t]\begin{center}
    \hfill{}
    \caption{The ID,  duration,  coordinates, and  redshift of the 361 GRBs
   as were published at http://lyra.berkeley.edu/grbox/grbox.php.}
        \begin{tabular}{|l||c|c|c|c|}\hline  
GRB  & T90  & RA  & Dec & z  \\ \hline \hline 
090429B  & 5.5 & 210.66688 & 32.17064 & 9.4 \\ \hline
090423 & 10.3 & 148.88871 & 18.14939 & 8.2 \\ \hline
120923A  & 27.2 & 303.79492 & 6.22119 & 8 \\ \hline
080913A  & 8 & 65.72775 & -25.1295 & 6.7 \\ \hline
060116 & 35 & 84.69283 & -5.43698 & 6.6 \\ \hline
050904 & 225 & 13.71221 & 14.08661 & 6.295 \\ \hline
120521C  & 26.7 & 214.28668 & 42.14478 & 6 \\ \hline
130606A  & 276.6 & 249.39662 & 29.7964 & 5.91 \\ \hline
060927 & 22.6 & 329.55008 & 5.36358 & 5.467 \\ \hline
050814 & 65 & 264.18912 & 46.33933 & 5.3 \\ \hline
071025 & 109 & 355.07116 & 31.77858 & 5.2 \\ \hline
050502B  & 7 & 142.54192 & 16.99625 & 5.2 \\ \hline
060522 & 69 & 322.93667 & 2.88621 & 5.11 \\ \hline
111008A  & 63.46 & 60.451 & -32.70928 & 4.9898 \\ \hline
060510B  & 276 & 239.12167 & 78.57 & 4.9 \\ \hline
100302A  & 17.9 & 195.51542 & 74.59014 & 4.813 \\ \hline
100513A  & 84 & 169.61129 & 3.62789 & 4.8 \\ \hline
100219A  & 18.8 & 154.20217 & -12.56656 & 4.6667 \\ \hline
090205 & 8.8 & 220.91104 & -27.85297 & 4.6497 \\ \hline
120401A  & 100 & 58.08258 & -17.63569 & 4.5 \\ \hline
000131 & 50 & 93.37917 & -51.94444 & 4.5 \\ \hline
060223A  & 11 & 55.2065 & -17.1301 & 4.41 \\ \hline
080916C  & 66 & 119.84717 & -56.63833 & 4.35 \\ \hline
080129 & 48 & 105.28404 & -7.84628 & 4.349 \\ \hline
050505 & 60 & 141.76392 & 30.2735 & 4.27 \\ \hline
120712A  & 14.7 & 169.58846 & -20.03383 & 4.1745 \\ \hline
090516A  & 210 & 138.26092 & -11.85428 & 4.109 \\ \hline
060206 & 11 & 202.93092 & 35.051 & 4.059 \\ \hline
100518A  & 30 & 304.78917 & -24.55456 & 4 \\ \hline
050730 & 155 & 212.07137 & -3.77158 & 3.9693 \\ \hline
120909A  & 112 & 275.73633 & -59.44836 & 3.93 \\ \hline
060210 & 255 & 57.73904 & 27.02622 & 3.9122 \\ \hline
090519 & 64 & 142.27917 & 0.18031 & 3.85 \\ \hline
081029 & 270 & 346.77233 & -68.15548 & 3.8479 \\ \hline
081228 & 3 & 39.46225 & 30.85292 & 3.8 \\ \hline
120802A  & 50 & 44.84313 & 13.76867 & 3.796 \\ \hline
050502A  & 20 & 202.44304 & 42.67425 & 3.793 \\ \hline
060605 & 15 & 322.1555 & -6.05869 & 3.773 \\ \hline
130408A  & 28 & 134.40542 & -32.36081 & 3.758 \\ \hline
60906 & 43.6 & 40.7535 & 39.36164 & 3.6856 \\ \hline
070721B  & 340 & 33.13729 & -2.19461 & 3.6298 \\ \hline
090323 & 150 & 190.70954 & 17.05322 & 3.57 \\ \hline
060115 & 142 & 54.03467 & 17.34531 & 3.5328 \\ \hline
0980329 & 15 & 105.65842 & 38.84556 & 3.5 \\ \hline
051028 & 12 & 27.06254 & 47.75256 & 3.5 \\ \hline
061110B  & 128 & 323.91833 & 6.87614 & 3.4344 \\ \hline
0971214 & 50 & 179.10979 & 65.21167 & 3.43 \\ \hline
060707 & 68 & 357.07917 & -17.90472 & 3.424 \\ \hline
121201A  & 85 & 13.46738 & -42.94289 & 3.385 \\ \hline
090313 & 78 & 198.40088 & 8.09717 & 3.375 \\ \hline
030323 & 25.05 & 166.53917 & -21.77033 & 3.372 \\ \hline
080810 & 106 & 356.79375 & 0.31944 & 3.3604 \\ \hline
        \end{tabular}
        \hfill{}
        \label{tab:err}
\end{center}
\end{table}

\begin{table}[t]\begin{center}
    \hfill{}
    \caption{ }
        \begin{tabular}{|l||c|c|c|c|}\hline  
GRB  & T90  & RA  & Dec & z  \\ \hline \hline 
110818A  & 103 & 317.33769 & -63.98119 & 3.36 \\ \hline
061222B  & 40 & 105.3525 & -25.86 & 3.355 \\ \hline
050908 & 20 & 20.46146 & -12.95478 & 3.3467 \\ \hline
050319 & 15 & 154.199 & 43.54858 & 3.2425 \\ \hline
060526 & 13.8 & 232.8265 & 0.2847 & 3.2213 \\ \hline
060926 & 8 & 263.93192 & 13.0385 & 3.2086 \\ \hline
080516 & 5.8 & 120.6415 & -26.15933 & 3.2 \\ \hline
020124 & 45.91 & 143.21171 & -11.51961 & 3.198 \\ \hline
100316A  & 7 & 251.97875 & 71.82708 & 3.155 \\ \hline
111123A  & 290 & 154.84642 & -20.64472 & 3.1516 \\ \hline
120922A  & 173 & 234.7485 & -20.18172 & 3.1 \\ \hline
091109A  & 48 & 309.25754 & -44.15822 & 3.076 \\ \hline
060607A  & 100 & 329.71 & -22.49631 & 3.0749 \\ \hline
081028A  & 260 & 121.89471 & 2.30808 & 3.038 \\ \hline
080607 & 79 & 194.94671 & 15.91969 & 3.0368 \\ \hline
121217A  & 778 & 153.71003 & -62.35098 & 3 \\ \hline
090404 & 84 & 239.23967 & 35.51597 & 3 \\ \hline
090715B  & 266 & 251.33967 & 44.83897 & 3 \\ \hline
070411 & 101 & 107.33304 & 1.06461 & 2.954 \\ \hline
120118B  & 23.26 & 124.871 & -7.18475 & 2.943 \\ \hline
051008 & 16 & 202.87313 & 42.09814 & 2.9 \\ \hline
060306 & 61 & 41.09546 & -2.14833 & 2.9 \\ \hline
050401 & 38 & 247.87008 & 2.18745 & 2.8983 \\ \hline
111107A  & 26.6 & 129.47775 & -66.52008 & 2.893 \\ \hline
120404A  & 38.7 & 235.0095 & 12.88503 & 2.876 \\ \hline
110731A  & 38.8 & 280.50412 & -28.53717 & 2.83 \\ \hline
050603 & 6 & 39.98705 & -25.18183 & 2.821 \\ \hline
120327A  & 62.9 & 246.86442 & -29.415 & 2.813 \\ \hline
130427B  & 27 & 314.89842 & -22.54636 & 2.78 \\ \hline
081222 & 24 & 22.73996 & -34.09486 & 2.77 \\ \hline
091029 & 39.2 & 60.17742 & -55.95556 & 2.752 \\ \hline
090809 & 5.4 & 328.67996 & -0.08384 & 2.737 \\ \hline
060714 & 115 & 227.86021 & -6.56619 & 2.7108 \\ \hline
090726 & 67 & 248.67935 & 72.88467 & 2.71 \\ \hline
121229A  & 100 & 190.10121 & -50.5943 & 2.707 \\ \hline
050406 & 3 & 34.46792 & -50.1875 & 2.7 \\ \hline
071031 & 180 & 6.40529 & -58.0595 & 2.6918 \\ \hline
080603B  & 60 & 176.53192 & 68.06111 & 2.6892 \\ \hline
120811C  & 26.8 & 199.68254 & 62.30075 & 2.671 \\ \hline
030429 & 9.19 & 183.28125 & -20.91381 & 2.6564 \\ \hline
080210 & 45 & 251.26671 & 13.82669 & 2.6419 \\ \hline
090529 & 100 & 212.469 & 24.45894 & 2.625 \\ \hline
070103 & 19 & 352.5575 & 26.87622 & 2.6208 \\ \hline
050215B  & 10 & 174.449 & 40.79581 & 2.62 \\ \hline
050820A  & 26 & 337.40879 & 19.56031 & 2.6147 \\ \hline
090426 & 1.2 & 189.07529 & 32.986 & 2.609 \\ \hline
130514A  & 204 & 296.28292 & -7.97622 & 2.6 \\ \hline
060923A  & 51.7 & 254.61733 & 12.36081 & 2.6 \\ \hline
080721 & 16.2 & 224.48273 & -11.72348 & 2.5914 \\ \hline
081118A  & 67 & 82.59242 & -43.30147 & 2.58 \\ \hline
050915A  & 25 & 81.68668 & -28.01646 & 2.5273 \\ \hline
081121 & 14 & 89.27579 & -60.60286 & 2.512 \\ \hline
050819 & 36 & 358.75675 & 24.86083 & 2.5043 \\ \hline
030115A  & 17.94 & 169.63596 & 15.04997 & 2.5 \\ \hline
070529 & 109 & 283.74246 & 20.65944 & 2.4996 \\ \hline
130518A  & 48 & 355.66781 & 47.46493 & 2.49 \\ \hline
120716A  & 230 & 313.05042 & 9.59825 & 2.48 \\ \hline
        \end{tabular}
        \hfill{}
        \label{tab:err}
\end{center}
\end{table}

\begin{table}[t]\begin{center}
    \hfill{}
    \caption{ }
        \begin{tabular}{|l||c|c|c|c|}\hline  
GRB  & T90  & RA  & Dec & z  \\ \hline \hline 
080515 & 21 & 3.1625 & 32.57806 & 2.47 \\ \hline
100424A  & 104 & 209.44762 & 1.53858 & 2.465 \\ \hline
070802 & 16.4 & 36.89867 & -55.52747 & 2.4541 \\ \hline
090812 & 66.7 & 353.20229 & -10.60472 & 2.452 \\ \hline
071021 & 225 & 340.64296 & 23.71847 & 2.452 \\ \hline
080413A  & 46 & 287.299 & -27.67781 & 2.433 \\ \hline
051001 & 190 & 350.95304 & -31.52314 & 2.4296 \\ \hline
080310 & 365 & 220.0575 & -0.17558 & 2.4274 \\ \hline
080905B  & 128 & 301.74121 & -62.56306 & 2.3739 \\ \hline
120815A  & 9.7 & 273.95758 & -52.13114 & 2.358 \\ \hline
070110 & 85 & 0.91363 & -52.97414 & 2.3521 \\ \hline
051109A  & 25 & 330.3138 & 40.82314 & 2.346 \\ \hline
110128A  & 30.7 & 193.89629 & 28.06544 & 2.339 \\ \hline
070129 & 460 & 37.00392 & 11.68444 & 2.3384 \\ \hline
021004 & 52.4 & 6.72783 & 18.92822 & 2.323 \\ \hline
060111A  & 13 & 276.205 & 37.60392 & 2.32 \\ \hline
070506 & 4.3 & 347.21829 & 10.72231 & 2.309 \\ \hline
121024A  & 69 & 70.47208 & -12.29069 & 2.298 \\ \hline
130505A  & 88 & 137.061 & 17.48478 & 2.27 \\ \hline
081221 & 34 & 15.79258 & -24.54769 & 2.26 \\ \hline
060124 & 710 & 77.10833 & 69.74089 & 2.23 \\ \hline
110205A  & 257 & 164.62967 & 67.52533 & 2.22 \\ \hline
080804 & 34 & 328.6675 & -53.18461 & 2.2045 \\ \hline
121128A  & 7.52 & 300.60004 & 54.29978 & 2.2 \\ \hline
050922C  & 5 & 317.38785 & -8.75839 & 2.1995 \\ \hline
070810A  & 11 & 189.96342 & 10.75119 & 2.17 \\ \hline
071020 & 4.2 & 119.66575 & 32.86111 & 2.1462 \\ \hline
011211 & 270 & 168.82492 & -21.94894 & 2.14 \\ \hline
060604 & 10 & 337.22921 & -10.9155 & 2.1357 \\ \hline
090926A  &  & 353.40015 & -66.32407 & 2.1062 \\ \hline
100728B  & 12.1 & 44.05617 & 0.28106 & 2.106 \\ \hline
060512 & 8.6 & 195.77421 & 41.1909 & 2.1 \\ \hline
081203A  & 294 & 233.03158 & 63.52081 & 2.1 \\ \hline
130610A  & 46.4 & 224.42033 & 28.20711 & 2.092 \\ \hline
061222A  & 72 & 358.26425 & 46.53294 & 2.088 \\ \hline
080207 & 340 & 207.51221 & 7.50186 & 2.0858 \\ \hline
000926 & 25 & 256.04046 & 51.78611 & 2.066 \\ \hline
070611 & 12 & 1.99171 & -29.75556 & 2.0394 \\ \hline
000301C  & 10 & 245.0775 & 29.44333 & 2.0335 \\ \hline
060108 & 14.4 & 147.00825 & 31.91906 & 2.03 \\ \hline
130612A  & 4 & 259.79408 & 16.71997 & 2.006 \\ \hline
121011A  & 75.6 & 260.21342 & 41.11039 & 2 \\ \hline
080906A  & 147 & 228.04438 & -80.51756 & 2 \\ \hline
030226 & 22.09 & 173.27054 & 25.89869 & 1.986 \\ \hline
081008 & 185.5 & 279.95833 & -57.43111 & 1.967 \\ \hline
070419B  & 236.5 & 315.70758 & -31.26369 & 1.9588 \\ \hline
050315 & 96 & 306.47542 & -42.60061 & 1.95 \\ \hline
080319C  & 34 & 258.98121 & 55.39183 & 1.9492 \\ \hline
060814 & 146 & 221.33871 & 20.58631 & 1.9229 \\ \hline
060708 & 9.8 & 7.80758 & -33.759 & 1.92 \\ \hline
020127 & 7.95 & 123.75592 & 36.77608 & 1.9 \\ \hline
060908 & 19.3 & 31.8265 & 0.342 & 1.8836 \\ \hline
131011A  & 77 & 32.52658 & -4.41119 & 1.874 \\ \hline
110801A  & 385 & 89.43721 & 80.95589 & 1.858 \\ \hline
090902B  & 21 & 264.93896 & 27.32419 & 1.822 \\ \hline
090709A  & 89 & 289.92767 & 60.72758 & 1.8 \\ \hline
120326A  & 69.6 & 273.90467 & 69.25986 & 1.798 \\ \hline
080325 & 128.4 & 277.89267 & 36.52342 & 1.78 \\ \hline
        \end{tabular}
        \hfill{}
        \label{tab:err}
\end{center}
\end{table}

\begin{table}[t]\begin{center}
    \hfill{}
    \caption{ }
        \begin{tabular}{|l||c|c|c|c|}\hline  
GRB  & T90  & RA  & Dec & z  \\ \hline \hline 
121027A  & 62.6 & 63.59767 & -58.82983 & 1.773 \\ \hline
110422 & 25.9 & 112.04608 & 75.10694 & 1.77 \\ \hline
100425A  & 37 & 299.1965 & -26.43081 & 1.755 \\ \hline
090113 & 9.1 & 32.0575 & 33.42842 & 1.7493 \\ \hline
120119A  & 253.8 & 120.02887 & -9.08158 & 1.728 \\ \hline
100906A  & 114.4 & 28.68379 & 55.63044 & 1.727 \\ \hline
050802 & 13 & 219.27371 & 27.78672 & 1.7102 \\ \hline
091020 & 34.6 & 175.72992 & 50.97831 & 1.71 \\ \hline
070521 & 37.9 & 242.66092 & 30.25622 & 1.7 \\ \hline
080928 & 280 & 95.07015 & -55.19971 & 1.6919 \\ \hline
080603A  & 180 & 279.40858 & 62.74425 & 1.688 \\ \hline
080605 & 20 & 262.12529 & 4.01556 & 1.6403 \\ \hline
0990510 & 100 & 204.53183 & -80.49689 & 1.619 \\ \hline
110503A  & 10 & 132.77608 & 52.20753 & 1.613 \\ \hline
0990123 & 63.3 & 231.37642 & 44.76642 & 1.61 \\ \hline
090418A  & 56 & 269.31321 & 33.40592 & 1.608 \\ \hline
071003 & 150 & 301.8505 & 10.94772 & 1.6044 \\ \hline
070714A  & 2 & 42.93046 & 30.24306 & 1.58 \\ \hline
100728A  & 198.5 & 88.75838 & -15.25567 & 1.567 \\ \hline
040912 & 150 & 359.179 & -0.92217 & 1.563 \\ \hline
051111 & 47 & 348.13783 & 18.37461 & 1.55 \\ \hline
070125 & 60 & 117.82403 & 31.15114 & 1.5471 \\ \hline
090102 &  & 128.24392 & 33.11419 & 1.547 \\ \hline
080520 & 2.8 & 280.19338 & -54.99197 & 1.5457 \\ \hline
060719 & 55 & 18.432 & -48.38092 & 1.532 \\ \hline
030328 & 92.59 & 182.70167 & -9.34758 & 1.522 \\ \hline
080330 & 61 & 169.26873 & 30.6232 & 1.5119 \\ \hline
080805 & 78 & 314.22267 & -62.44439 & 1.5042 \\ \hline
060502A  & 33 & 240.927 & 66.60069 & 1.5026 \\ \hline
070306 & 210 & 148.09713 & 10.48202 & 1.49594 \\ \hline
060418 & 52 & 236.4275 & -3.63889 & 1.49 \\ \hline
120724A  & 72.8 & 245.18062 & 3.50772 & 1.48 \\ \hline
010222 &  & 223.05229 & 43.01839 & 1.478 \\ \hline
110213A  & 48 & 42.96429 & 49.27314 & 1.46 \\ \hline
090407 & 310 & 68.97975 & -12.67922 & 1.4485 \\ \hline
050318 & 32 & 49.71312 & -46.39547 & 1.4436 \\ \hline
100814A  & 174.5 & 22.47338 & -17.99544 & 1.44 \\ \hline
050822 & 102 & 51.11342 & -46.03333 & 1.434 \\ \hline
080604 & 82 & 236.96542 & 20.55781 & 1.4171 \\ \hline
100901A  & 439 & 27.26425 & 22.75856 & 1.408 \\ \hline
120711A  & 44 & 94.6785 & -70.99911 & 1.405 \\ \hline
080602 & 74 & 19.17571 & -9.23219 & 1.4 \\ \hline
100615A  & 39 & 177.20542 & -19.48117 & 1.398 \\ \hline
111229A  & 25.4 & 76.28692 & -84.71086 & 1.3805 \\ \hline
050801 & 20 & 204.14583 & -21.92806 & 1.38 \\ \hline
090927 & 2.2 & 343.97254 & -70.98036 & 1.37 \\ \hline
100414A  & 26.4 & 192.11233 & 8.69303 & 1.368 \\ \hline
110808A  & 48 & 57.26783 & -44.19453 & 1.348 \\ \hline
071117 & 6.6 & 335.04342 & -63.44319 & 1.331 \\ \hline
061121 & 81 & 147.22742 & -13.1952 & 1.3145 \\ \hline
0990506 & 150 & 178.70892 & -26.67644 & 1.307 \\ \hline
130511A  & 5.43 & 196.64567 & 18.71 & 1.3033 \\ \hline
130420A  & 123.5 & 196.10654 & 59.42408 & 1.297 \\ \hline
050126 & 26 & 278.11321 & 42.37044 & 1.29 \\ \hline
100724A  & 1.4 & 194.54333 & -11.1025 & 1.288 \\ \hline
061007 & 75 & 46.33167 & -50.50069 & 1.2622 \\ \hline
020813A  & 88.98 & 296.67446 & -19.60134 & 1.2545 \\ \hline
090926B  & 81 & 46.30808 & -39.00617 & 1.24 \\ \hline
        \end{tabular}
        \hfill{}
        \label{tab:err}
\end{center}
\end{table}

\begin{table}[t]\begin{center}
    \hfill{}
    \caption{ }
        \begin{tabular}{|l||c|c|c|c|}\hline  
GRB  & T90  & RA  & Dec & z  \\ \hline \hline 
130907A  & 115.1 & 215.892 & 45.60742 & 1.238 \\ \hline
050408 & 34 & 180.57212 & 10.85261 & 1.2356 \\ \hline
080707 & 27.1 & 32.61833 & 33.10953 & 1.2322 \\ \hline
130418A  & 300 & 149.03717 & 13.66744 & 1.218 \\ \hline
100316B  & 3.8 & 163.48812 & -45.47267 & 1.18 \\ \hline
060319 & 12 & 176.38704 & 60.01086 & 1.172 \\ \hline
070208 & 48 & 197.88586 & 61.9651 & 1.165 \\ \hline
070518 & 5.5 & 254.19875 & 55.29508 & 1.16 \\ \hline
061126 & 191 & 86.60198 & 64.21068 & 1.159 \\ \hline
130701A  & 4.38 & 357.22954 & 36.10039 & 1.155 \\ \hline
071122 & 68.7 & 276.60525 & 47.07514 & 1.14 \\ \hline
060801 & 0.5 & 213.00554 & 16.98183 & 1.131 \\ \hline
000418 & 30 & 186.33042 & 20.10322 & 1.11854 \\ \hline
0981226 & 260 & 352.40417 & 22.93161 & 1.11 \\ \hline
080413B  & 8 & 326.1445 & -19.98111 & 1.1014 \\ \hline
0980613 & 50 & 154.49092 & 71.45708 & 1.0964 \\ \hline
091024 & 1200 & 339.24875 & 56.88983 & 1.092 \\ \hline
110213B  &  & 41.75588 & 1.14619 & 1.083 \\ \hline
091208B  & 14.9 & 29.39204 & 16.88967 & 1.0633 \\ \hline
051006 & 26 & 110.80633 & 9.5068 & 1.059 \\ \hline
000911 & 500 & 34.64317 & 7.74103 & 1.0585 \\ \hline
110726A  & 5.2 & 286.71692 & 56.07128 & 1.036 \\ \hline
080411 & 56 & 37.97996 & -71.30203 & 1.0301 \\ \hline
121211A  & 182 & 195.53329 & 30.1485 & 1.023 \\ \hline
0991216 & 50 & 77.38041 & 11.28535 & 1.02 \\ \hline
021211 & 2.8 & 122.24951 & 6.72719 & 1.006 \\ \hline
110918A  & 22 & 32.53912 & -27.10544 & 0.982 \\ \hline
071010A  & 6 & 288.06093 & -32.40199 & 0.98 \\ \hline
081109 & 190 & 330.7905 & -54.71097 & 0.9787 \\ \hline
091018 & 4.4 & 32.18588 & -57.54828 & 0.971 \\ \hline
120907A  & 16.9 & 74.75 & -9.315 & 0.97 \\ \hline
070419A  & 116 & 182.74517 & 39.92533 & 0.97 \\ \hline
0980703 & 40 & 359.77779 & 8.5853 & 0.967 \\ \hline
120722A  & 42.4 & 230.4966 & 13.2513 & 0.9586 \\ \hline
0970828 & 160 & 272.10629 & 59.30236 & 0.958 \\ \hline
071010B  & 35.7 & 150.53858 & 45.73064 & 0.947 \\ \hline
071028B  &  & 354.16167 & -31.62047 & 0.94 \\ \hline
080319B  & 50 & 217.92075 & 36.30244 & 0.9382 \\ \hline
060912A  & 5 & 5.284 & 20.97161 & 0.937 \\ \hline
051016B  & 4 & 132.11583 & 13.65575 & 0.9364 \\ \hline
070714B  & 64 & 57.8425 & 28.29761 & 0.923 \\ \hline
090510 & 0.3 & 333.55267 & -26.58411 & 0.903 \\ \hline
070429B  & 0.5 & 328.01587 & -38.82833 & 0.9023 \\ \hline
091003 & 21.1 & 251.51953 & 36.62521 & 0.8969 \\ \hline
040924 & 1.2 & 31.594 & 16.11344 & 0.859 \\ \hline
101225A  & 1088 & 0.19792 & 44.60067 & 0.847 \\ \hline
080710 & 120 & 8.27354 & 19.50147 & 0.8454 \\ \hline
000210 & 20 & 29.81496 & -40.65917 & 0.8452 \\ \hline
0990705 & 45 & 77.47708 & -72.13139 & 0.842 \\ \hline
070318 & 63 & 48.48679 & -42.94619 & 0.84 \\ \hline
0970508 & 35 & 103.45604 & 79.27208 & 0.835 \\ \hline
050824 & 25 & 12.23421 & 22.60922 & 0.8278 \\ \hline
061217 & 0.3 & 160.41383 & -21.12281 & 0.827 \\ \hline
071112C  & 15 & 39.21221 & 28.37131 & 0.8227 \\ \hline
110715A  & 13 & 237.68371 & -46.23515 & 0.82 \\ \hline
070508 & 21 & 312.80029 & -78.38528 & 0.82 \\ \hline
051022 & 200 & 359.01708 & 19.60669 & 0.809 \\ \hline
100816A  & 2.9 & 351.73983 & 26.57858 & 0.804 \\ \hline
        \end{tabular}
        \hfill{}
        \label{tab:err}
\end{center}
\end{table}

\begin{table}[t]\begin{center}
    \hfill{}
    \caption{ }
        \begin{tabular}{|l||c|c|c|c|}\hline  
GRB  & T90  & RA  & Dec & z  \\ \hline \hline 
120729A  & 71.5 & 13.07429 & 49.93975 & 0.8 \\ \hline
060602A  & 60 & 149.56938 & 0.30408 & 0.787 \\ \hline
060202 & 203.7 & 35.84587 & 38.38422 & 0.785 \\ \hline
030528 & 53.85 & 256.00129 & -22.61944 & 0.782 \\ \hline
080430 & 16.2 & 165.31129 & 51.68569 & 0.767 \\ \hline
061110A  & 41 & 336.29146 & -2.25886 & 0.7578 \\ \hline
090328 & 80 & 90.66529 & -41.88161 & 0.736 \\ \hline
050813 & 0.6 & 241.98737 & 11.24919 & 0.72 \\ \hline
101219A  & 0.6 & 74.58537 & -2.53972 & 0.718 \\ \hline
131004A  & 1.54 & 296.11283 & -2.95839 & 0.717 \\ \hline
041006 &  & 13.70929 & 1.23469 & 0.716 \\ \hline
111228A  & 101.2 & 150.06671 & 18.29772 & 0.716 \\ \hline
051227 & 8 & 125.24212 & 31.92553 & 0.714 \\ \hline
0991208 & 60 & 248.473 & 46.45583 & 0.7055 \\ \hline
060904B  & 192 & 58.2105 & -0.72525 & 0.7029 \\ \hline
090814A  & 80 & 239.60979 & 25.63122 & 0.696 \\ \hline
0970228 & 3.6 & 75.44421 & 11.7815 & 0.695 \\ \hline
020405 & 40 & 209.513 & -31.37275 & 0.695 \\ \hline
080916A  & 60 & 336.27579 & -57.023 & 0.6887 \\ \hline
111209A  & 1400 & 14.34492 & -46.80117 & 0.677 \\ \hline
050416A  & 2.4 & 188.47747 & 21.0573 & 0.6528 \\ \hline
100418A  & 7 & 256.36287 & 11.46175 & 0.6235 \\ \hline
110106B  & 24.8 & 134.15528 & 47.00291 & 0.618 \\ \hline
070612A  & 370 & 121.37337 & 37.27089 & 0.617 \\ \hline
050525A  & 10 & 278.13571 & 26.33958 & 0.606 \\ \hline
130215A  & 65.7 & 43.50292 & 13.39539 & 0.597 \\ \hline
050223 & 23 & 271.38538 & -62.47252 & 0.5915 \\ \hline
060123 & 900 & 179.69933 & 45.51394 & 0.56 \\ \hline
101219B  & 34 & 12.23063 & -34.56653 & 0.5519 \\ \hline
051221A  & 1.4 & 328.70261 & 16.89088 & 0.5465 \\ \hline
090424 & 52 & 189.52129 & 16.83753 & 0.544 \\ \hline
060729 & 116 & 95.38246 & -62.37022 & 0.5428 \\ \hline
100621A  & 63.6 & 315.3045 & -51.10625 & 0.542 \\ \hline
090618 & 113.2 & 293.9955 & 78.35686 & 0.54 \\ \hline
081007A  & 10 & 339.96 & -40.14689 & 0.5295 \\ \hline
091127 & 7.1 & 36.58288 & -18.95236 & 0.49 \\ \hline
051117B  & 8 & 85.18075 & -19.27422 & 0.481 \\ \hline
130831A  & 32.5 & 358.62458 & 29.42967 & 0.4791 \\ \hline
111211A  & 15 & 153.09042 & 11.20833 & 0.478 \\ \hline
070724A  & 0.4 & 27.80863 & -18.59426 & 0.457 \\ \hline
010921 & 21.77 & 343.99958 & 40.93139 & 0.45 \\ \hline
061006 & 130 & 111.03192 & -79.19864 & 0.4377 \\ \hline
0990712 & 30 & 337.97096 & -73.40786 & 0.43 \\ \hline
020819 & 20 & 351.83112 & 6.26554 & 0.41 \\ \hline
061210 & 85 & 144.52196 & 15.62147 & 0.4095 \\ \hline
120714B  & 159 & 355.40875 & -46.18389 & 0.3984 \\ \hline
071227 & 1.8 & 58.13025 & -55.98431 & 0.394 \\ \hline
011121 &  & 173.62346 & -76.02819 & 0.362 \\ \hline
130603B  & 0.18 & 172.20063 & 17.07167 & 0.356 \\ \hline
110328A  &  & 251.20805 & 57.58325 & 0.354 \\ \hline
060428B  & 58 & 235.35679 & 62.02508 & 0.35 \\ \hline
130925A  & 0 & 41.179 & -26.1531 & 0.347 \\ \hline
        \end{tabular}
        \hfill{}
        \label{tab:err}
\end{center}
\end{table}

\begin{table}[t]\begin{center}
    \hfill{}
    \caption{ }
        \begin{tabular}{|l||c|c|c|c|}\hline  
GRB  & T90  & RA  & Dec & z  \\ \hline \hline 
061021 & 46 & 145.15058 & -21.95122 & 0.3463 \\ \hline
090417B  & 260 & 209.69412 & 47.01806 & 0.345 \\ \hline
130427A  & 162.8 & 173.13683 & 27.69894 & 0.34 \\ \hline
050826 & 35 & 87.75658 & -2.64328 & 0.296 \\ \hline
060502B  & 90 & 278.93971 & 52.63136 & 0.287 \\ \hline
120422A  & 5.35 & 136.90992 & 14.01875 & 0.283 \\ \hline
050724 & 3 & 246.18487 & -27.54097 & 0.258 \\ \hline
020903 & 32.15 & 342.17642 & -20.76925 & 0.251 \\ \hline
050509B  & 0.13 & 189.05858 & 28.98533 & 0.2249 \\ \hline
070809 & 1.3 & 203.76896 & -22.14189 & 0.2187 \\ \hline
081211B  & 102 & 168.26404 & 53.82992 & 0.216 \\ \hline
040701 & 60 & 312.06708 & -40.18579 & 0.2146 \\ \hline
030329A  & 22.76 & 161.20817 & 21.52151 & 0.1687 \\ \hline
050709 & 220 & 345.36233 & -38.97764 & 0.16 \\ \hline
130702A  & 59 & 217.31158 & 15.774 & 0.145 \\ \hline
000607 & 0.15 & 38.49475 & 17.14764 & 0.1405 \\ \hline
060614 & 102 & 320.88367 & -53.02672 & 0.1257 \\ \hline
061201 & 0.8 & 332.13371 & -74.57974 & 0.111 \\ \hline
031203 & 30 & 120.6265 & -39.85003 & 0.105 \\ \hline
060505 & 4 & 331.76433 & -27.81442 & 0.089 \\ \hline
051109B  & 15 & 345.45983 & 38.67964 & 0.08 \\ \hline
100316D  & 240 & 107.62642 & -56.25464 & 0.059 \\ \hline
060218 & 2100 & 50.41535 & 16.86717 & 0.0331 \\ \hline
111005A  & 26 & 223.28242 & -19.73672 & 0.01326 \\ \hline
0980425 & 30 & 293.76379 & -52.84575 & 0.0085 \\ \hline
080109 & 500 & 137.37771 & 33.13897 & 0.006494 \\ \hline
        \end{tabular}
        \hfill{}
        \label{tab:err}
\end{center}
\end{table}

\end{document}